\documentclass[prd,twocolumn]{revtex4}\begin{document}
\title{Power spectrum of stochastic wave and  diffusion equations in the warm inflation models}
\author{ Z. Haba\\
Institute of Theoretical Physics, University of Wroclaw,\\ 50-204
Wroclaw, Plac Maxa Borna 9,
Poland}\email{zbigniew.haba@uwr.edu.pl}
\begin{abstract} We discuss dissipative
stochastic wave and diffusion equations resulting from an
interaction of the inflaton with an environment in an external
expanding homogeneous metric. We show that a diffusion equation
well approximates the wave equation in a strong friction limit. We
calculate  the long wave power spectrum of the wave equation under
the assumption that the perturbations are slowly varying in time
and the expansion is almost exponential. Under the assumption that
the noise has a form invariant under the coordinate
transformations we obtain the power spectrum close to the scale
invariant one. In the diffusion approximation we go beyond the
slow variation assumption.  We calculate the power spectrum
exactly in models with exponential inflation and polynomial
potentials and with power-law inflation and exponential
potentials.
\end{abstract} \maketitle
\section{Introduction}It has been discovered
long time ago by Harrison \cite{har} and Zeldovich \cite{zel} that
the scale invariant spectrum of galaxy distribution plays an
essential role in the  galaxy formation after the Big Bang. The
spectrum close to the scale invariant one has been confirmed by
WMAP observations \cite{sto}\cite{ade}. Since then the power
spectrum is a substantial criterion for a validity of cosmological
models.
 The power spectrum
results from quantization of the quadratic fluctuations around the
homogeneous solution
\cite{starjetpl}\cite{muk}\cite{starjpl}\cite{lin1}\cite{vil}\cite{guthpi}\cite{linde}.
The formalism is explicitly gauge invariant
\cite{bardeen}\cite{feld}. It treats quantum gravitational
fluctuations and inflaton fluctuations on the same footing.
Nevertheless, it has been shown
\cite{starjettpl}\cite{marozzi}\cite{marozzi2}\cite{starven} that
if the quantum inflaton fluctuations are expressed by stochastic
fluctuations in e-fold time then the inflaton fluctuations already
contain the gravitational fluctuations leading to the same power
spectrum as in \cite{feld}. The scalar field models can be
considered as effective field theories of scalar cosmological
perturbations. The power spectrum close to the scale invariant one
has also been obtained in warm inflation models
\cite{hal}\cite{ram}.  The model of warm inflation is treated as
an effective field theory of an inflaton interacting with a large
number of other fields \cite{berera2}\cite{bererarev}. The
inflaton wave equation becomes stochastic as a result of an
interaction with other fields whose effect is described on the the
basis of their contribution to the entropy and to the energy
density \cite{hal}\cite{mosstaylor}\cite{graham}.
 In this paper we
study in detail the stochastic wave equation in the form derived
from an interaction of scalar fields with an environment
\cite{ber1} \cite{hab1}\cite{habaepj}. In this model  the heat
bath is an initial state of an infinite set of scalar massive
fields $\chi_{n}$ interacting with the inflaton. The fields
$\chi_{n}$ are treated as unobservable degrees of freedom. We
average over these degrees of freedom. As a result of the
interaction with $\chi_{n}$  the inflaton acquires a friction and
a noise term. The stochastic model is considered as a Markovian
approximation to the Hamiltonian model of the $\chi_{n}$ fields
interacting with the inflaton. Such an approach is analogous to
the treatment of a Brownian particle in an environment of other
particles.

In this paper we first  repeat a calculation of the power spectrum
in an extended model of the inflaton-environment interaction
\cite{habaepj} on the basis of previously developed methods
\cite{bass}\cite{habapl} under the assumption that the
perturbations of the non-linear wave equation are slowly varying
in time (are almost constant). However, the main objective here
is a  development of  another tool for a computation of the power
spectrum based on the approximation of the dissipative wave
equation by a diffusion equation.
  In the case of a random diffusion the calculation of the power
spectrum is much simpler. Its dependence on the evolution law can
be seen in a more transparent way. We are able to calculate the
power spectrum with parameters which can substantially vary in
time.

The plan of the paper is the following. In sec.2 we repeat the
main steps of the derivation \cite{habaepj} of the model
emphasizing the extra terms which appear in comparison to the warm
inflation inflaton equation \cite{bererarev}. We discuss the
resulting dissipative stochastic wave equation for an inflaton
interacting with an environment by means of  a potential $U$ . In
sec.3 we show that at strong friction the solutions of the wave
equation tend to the solutions of a diffusion
 equation. In sec.4 we calculate the
power spectrum of the stochastic wave equation under the
assumption that the evolution of the scale factor is almost
exponential and  the variables in this equation can be treated as
constants. This is a repetition of the standard calculations
\cite{bass}\cite{habapl}\cite{ram} but in a model with different
potentials and a different noise. In sec.5 we calculate the power
spectrum of the stochastic diffusion equation assuming again that
the expansion of the metric is almost exponential and that the
variables in this equation can be treated as  constants. We obtain
the same power spectrum as in the case of the wave equation,i.e.,
an almost scale invariant spectrum, which is shown to be a
consequence of the form of the noise. We discuss a relation of our
results to the ones in the literature.In sec.6 we study solutions
of the diffusion equation with almost exponential expansion but
with varying potentials. We obtain a shift in the formula for the
spectral index. Then, we explore  exponential potentials in a
power-law expanding metric when the method of constant parameters
does not apply. The diffusion equation reveals a sensitive
dependence of the power spectrum on the potentials. In Appendix A
we show in a simple way that the form of the noise that leads to
the scale invariant spectrum follows from its invariance under
coordinate transformations. In Appendix B we give a simplified
derivation of the scale invariant spectrum showing the crucial
role of the form of the  noise and the exponential expansion.

\section{The model of an environment }
We recall  the basic ingredients of the model
\cite{ber1}\cite{hab1}\cite{habaepj} of an interaction of the
inflaton with an environment. We consider the Lagrangian
\begin{equation}\begin{array}{l}
{\cal L}=\frac{1}{2}\partial_{\mu}\phi\partial^{\mu}\phi
-V(\phi)\cr+\sum_{n}(\frac{1}{2}\partial_{\mu}\chi_{n}\partial^{\mu}\chi_{n}
-\frac{1}{2}m_{n}^{2}\chi_{n}\chi_{n}-\lambda_{n}U(\phi)\chi_{n}),
\end{array}\end{equation}where $U(\phi)$ is a certain interaction potential. Equations of motion read
\begin{equation}
g^{-\frac{1}{2}}\partial_{\mu}(g^{\frac{1}{2}}\partial^{\mu}\phi)=-V^{\prime}-U^{\prime}(\phi)\sum_{n}\lambda_{n}\chi_{n},
\end{equation}
\begin{equation}
g^{-\frac{1}{2}}\partial_{\mu}(g^{\frac{1}{2}}\partial^{\mu}\chi_{n})+m_{n}^{2}\chi_{n}=-\lambda_{n}U(\phi),
\end{equation}where $g_{\mu\nu}$ is the metric tensor and $g=\vert \det[g_{\mu\nu}]\vert$.
We restrict ourselves to the flat expanding metric
\begin{equation}
ds^{2}=g_{\mu\nu}dx^{\mu}dx^{\nu}=dt^{2}-a^{2}d{\bf x}^{2}
\end{equation}
We write
\begin{equation}
\chi_{n}=a^{-\frac{3}{2}}\tilde{\chi}_{n}.
\end{equation}
Then, in the momentum space eq.(3) reads\begin{equation}
\partial_{t}^{2}\tilde{\chi}_{n}+\omega_{n}^{2}\tilde{\chi}_{n}=-\lambda_{n}a^{\frac{3}{2}}U(\phi),
\end{equation} where
\begin{displaymath}
\omega_{n}^{2}=a^{-2}k^{2}+m_{n}^{2}-\frac{3}{2}\partial_{t}H-\frac{9}{4}H^{2}
\end{displaymath}  $H=a^{-1}\partial_{t}a$. Let us consider low momenta $a^{-1}k<<H$ so that for a large time
we can neglect $a^{-2}k^{2} $ term. We also assume that $H$ is
slowly varying and $\omega_{n}^{2}>0$.

We can solve eq.(3) for $\chi_{n}$
\begin{equation}
\chi_{n}=\chi_{n}^{cl}-\lambda_{n}\int dx^{\prime}
G_{n}(x,x^{\prime})U(x^{\prime}),\end{equation} where we denote
$U(x)=U(\phi(x))$, $\chi_{n}^{cl}$ are solutions of the linear
equation and $G_{n}$ is the Green function. When we insert
$\chi_{n}$ of eq.(7) in eq.(2) then it takes the form
\begin{equation}\begin{array}{l}
g^{-\frac{1}{2}}\partial_{\mu}(g^{\frac{1}{2}}\partial^{\mu}\phi)+V^{\prime}=-U^{\prime}\sum_{n}\lambda_{n}^{2}G_{n}
U  +U^{\prime}\eta\equiv \delta\phi +U^{\prime}\tilde{\eta},
\end{array}\end{equation} where
\begin{equation}
\tilde{\eta}=-\sum_{n}\lambda_{n}\chi^{cl}_{n}.
\end{equation}
When we calculate the expectation value of $\tilde{\eta}$ over the
free field solutions $\chi_{n}^{cl}$ with respect to the Gibbs
measure (with temperature $\beta^{-1}$) then approximately
\begin{equation}\begin{array}{l}
\langle\tilde{\eta}(x)\tilde{\eta}(x^{\prime})\rangle
=\beta^{-1}(2\pi)^{-3}\int d{\bf k} \exp(i{\bf k}({\bf x}-{\bf
x}^{\prime}))\cr\sum_{n}\lambda_{n}^{2}\omega_{n}^{-2}
\cos(\omega_{n}(t-t^{\prime})) .\end{array}\end{equation} We
assume that  $\lambda_{n}^{2}\omega_{n}^{-2}\simeq \gamma^{2}$ is
 a constant. Then
\begin{equation}
\langle\tilde{\eta}(x)\tilde{\eta}(x^{\prime})\rangle
=\beta^{-1}\gamma^{2}a^{-3}\delta({\bf x}-{\bf
x}^{\prime})\delta(t-t^{\prime}).
\end{equation}
The friction $\delta \phi$ in eq.(8) can be expressed by means of
an approximate Green function
\begin{displaymath}
\frac{\sin(\omega_{n}(t-s))}{\omega_{n}}=\partial_{s}\Big(\frac{\cos(\omega_{n}(t-s))}{\omega_{n}^{2}}\Big)
\end{displaymath}
(in eq.(7) $x=(t,{\bf x})$ and $x^{\prime}=(s,{\bf x}^{\prime})$).
 After an
integration by parts in $G_{n}U$ the friction term takes the form
\begin{equation}
-a(t)^{-\frac{3}{2}}U^{\prime}\sum_{n}\lambda_{n}^{2}\int
\omega_{n}^{-2}\cos(\omega_{n}(t-s))\partial_{s}(a(s)^{\frac{3}{2}}U)ds.
\end{equation}
We have got the same kernel as the one in eq.(10) which we
approximated by $\delta(t-s)$ in eq.(11) for the correlation of
the noise.  With these approximations, when we perform the
differentiation in eq.(12), then eq.(2)in the flat expanding
metric (4) reads ( we change the notation
$\phi\rightarrow\phi_{\eta}$)
\begin{equation}\begin{array}{l}
\partial_{t}^{2}\phi_{\eta}-a^{-2}\triangle\phi_{\eta}+(3H+\gamma^{2}(U^{\prime})^{2})\partial_{t}\phi_{\eta}
\cr+V^{\prime}(\phi_{\eta})
+\frac{3}{2}\gamma^{2}HUU^{\prime}(\phi_{\eta})=\beta^{-\frac{1}{2}}\gamma
a^{-\frac{3}{2}}U^{\prime}\eta,
 \end{array}\end{equation}where
we wrote
$\tilde{\eta}=\gamma\beta^{-\frac{1}{2}}a^{-\frac{3}{2}}\eta$ so
that
\begin{equation}
\langle \eta_{s}({\bf x})\eta_{t}({\bf
y})\rangle=\delta(t-s)\delta({\bf x}-{\bf y}).
\end{equation}

We consider a linearized form of eq.(13) resulting from an
expansion around its homogeneous (space-independent)
solution\begin{equation}
\partial_{t}^{2}\phi_{c}+(3H+\gamma^{2}U^{\prime}(\phi_{c})^{2})\partial_{t}\phi_{c}+V^{\prime}(\phi_{c})
+\frac{3}{2}\gamma^{2}HUU^{\prime}(\phi_{c})=0
\end{equation}
 We
write  $\phi_{\eta}=\phi_{c}+\phi$. The initial conditions are
contained in $\phi_{c}$, so we assume zero as the intitial
condition for $\phi$. The linearization of eq.(13) expanded about
$\phi_{c}$ reads
\begin{equation}\begin{array}{l}
\partial_{t}^{2}\phi-a^{-2}\triangle\phi+(3H+\gamma^{2}U^{\prime}(\phi_{c})^{2})\partial_{t}\phi+V^{\prime\prime}(\phi_{c})\phi
\cr
+2\gamma^{2}U^{\prime}U^{\prime\prime}(\phi_{c})\partial_{t}\phi_{c}\phi
+\frac{3}{2}\gamma^{2}H((U^{\prime})^{2}+UU^{\prime\prime})\phi\cr=\beta^{-\frac{1}{2}}\gamma
a^{-\frac{3}{2}}U^{\prime}(\phi_{c})\eta.
 \end{array}\end{equation} We can transform eq.(16) to another form . Let
\begin{equation}
\phi=a^{-\frac{3}{2}}\exp\Big(-\frac{1}{2}\gamma^{2}\int_{0}^{t}U^{\prime}(\phi_{c})^{2}\Big)\Phi.
\end{equation}
Then

\begin{equation}\begin{array}{l}
\partial_{t}^{2}\Phi-a^{-2}\triangle\Phi-\Omega^{2}\Phi\cr=\beta^{-\frac{1}{2}}\gamma
\exp\Big(\frac{1}{2}\gamma^{2}\int_{0}^{t}U^{\prime}(\phi_{c})^{2}\Big)U^{\prime}(\phi_{c})\eta_{t},
 \end{array}\end{equation}
where
\begin{displaymath}\begin{array}{l}
\Omega^{2}=-V^{\prime\prime}-\frac{1}{2}\gamma^{2}\partial_{t}(U^{\prime})^{2}
+\frac{3}{2}\partial_{t}H
\cr-\frac{3}{2}\gamma^{2}H((U^{\prime})^{2}+UU^{\prime\prime})+\frac{1}{4}(3H+\gamma^{2}
(U^{\prime})^{2})^{2} .
\end{array}\end{displaymath}The wave equation with friction is
transformed into a wave equation with a  complex mass $\Omega$.
Note that large $3H+\gamma^{2}(U^{\prime})^{2}$ means large
$\Omega$.

\section{Diffusion approximation}
In this section we show that the diffusion approximation to
eq.(16) , i.e., the omission of $\partial_{t}^{2}\phi$,  is
equivalent to the neglect of fast decaying modes (for  a large
$\Omega$) in the solution of eqs.(17)-(18). The diffusion
approximation to eq.(16) in the momentum space reads (we denote
the Fourier transform $\tilde{\phi}({\bf k})$ by the same letter
as its spatial form $\phi({\bf
x})$)\begin{equation}\begin{array}{l}
(3H+\gamma^{2}U^{\prime}(\phi_{c})^{2})\partial_{t}\phi+a^{-2}k^{2}\phi++V^{\prime\prime}(\phi_{c})\phi
\cr
+2\gamma^{2}U^{\prime}U^{\prime\prime}(\phi_{c})\partial_{t}\phi_{c}\phi
\cr
+\frac{3}{2}\gamma^{2}H((U^{\prime})^{2}+UU^{\prime\prime})\phi=\beta^{-\frac{1}{2}}\gamma
a^{-\frac{3}{2}}U^{\prime}(\phi_{c})\eta.\end{array}
\end{equation}
On the other hand we may express the solution of eq.(18)(momentum
space) with zero initial condition at $t_{0}$ by means of the
Green function $G$
\begin{equation}
\Phi(t)=\beta^{-\frac{1}{2}}\gamma\int_{t_{0}}^{t}
G(t,s)\exp\Big(\frac{1}{2}\gamma^{2}\int_{0}^{s}U^{\prime}(\phi_{c})^{2}\Big)U^{\prime}(\phi_{c})\eta_{s}ds,\end{equation}where
the approximate Green function (for large slowly varying $\omega$)
is
\begin{equation}
G(t,s)=\omega(s)^{-\frac{1}{2}}\omega(t)^{-\frac{1}{2}}\sinh\Big(\int_{s}^{t}d\tau\omega(\tau)\Big)
\end{equation}with
\begin{equation}
\omega^{2}=\Omega^{2}-a^{-2}{\bf k}^{2}.\end{equation} Expanding
$\omega$ in powers of $(3H+\tilde{\gamma}^{2})^{-1}$ where
\begin{equation}
\tilde{\gamma}^{2}=\gamma^{2}(U^{\prime})^{2}
\end{equation}
we obtain in the lowest order of the expansion
\begin{equation}\begin{array}{l}
\omega=\frac{3}{2}H+\frac{1}{2}\tilde{\gamma}^{2}+(3H+\tilde{\gamma}^{2})^{-1}\Big
(-a^{-2}k^{2}-V^{\prime\prime}\cr-2\gamma^{2}U^{\prime}U^{\prime\prime}\partial_{t}\phi_{c}
\cr-\frac{3}{2}\gamma^{2}H((U^{\prime})^{2}+UU^{\prime\prime})+\frac{3}{2}\partial_{t}H+
\frac{1}{2}\gamma^{2}\partial_{t}(U^{\prime})^{2}\Big) \cr\equiv
\frac{3}{2}H+\frac{1}{2}\tilde{\gamma}^{2}+\frac{1}{2}\partial_{t}\ln(3H+\tilde{\gamma}^{2})
 -v,
\end{array}\end{equation} where
\begin{equation}\begin{array}{l}
v=(3H+\tilde{\gamma}^{2})^{-1}\Big(a^{-2}k^{2}+V^{\prime\prime}+\frac{1}{2}\partial_{t}\tilde{\gamma}^{2}
\cr+\frac{3}{2}\gamma^{2}H((U^{\prime})^{2}+UU^{\prime\prime})
\Big).
\end{array}\end{equation}
We compare solutions of the wave equation (16) with solutions of
the diffusion equation (19). The solution of the diffusion
equation (19) is
\begin{equation}\begin{array}{l}
\phi_{t}=\beta^{-\frac{1}{2}}\gamma
\int_{t_{0}}^{t}\exp\Big(-\int_{s}^{t} d\tau
v(\tau)\Big)\cr(3H(s)+\tilde{\gamma}^{2})^{-1}a(s)^{-\frac{3}{2}}U^{\prime}(\phi_{c})\eta(s)ds.
\end{array}\end{equation} We compare the solution (26) with (20)-(21). In
the Green function (21) we have
\begin{equation}\begin{array}{l}
\int_{s}^{t}d\tau\omega(\tau)=\frac{1}{2}\int_{s}^{t}(3H(\tau)+\tilde{\gamma}(\tau)^{2})d\tau
\cr+\frac{1}{2}\ln(3H(t)+\tilde{\gamma}(t)^{2})-\frac{1}{2}\ln(3H(s)+\tilde{\gamma}(s)^{2})-\int_{s}^{t}d\tau
v(\tau).
\end{array}\end{equation}
If in
\begin{displaymath}
\sinh(X)=\frac{1}{2}\exp(X)-\frac{1}{2}\exp(-X)
\end{displaymath}
we neglect the second term as quickly vanishing (for $X>0$) and in
eq.(21) $\omega(s)^{-\frac{1}{2}}$ is approximated by
$(\frac{3}{2}H+\frac{1}{2}\tilde{\gamma}^{2})^{-\frac{1}{2}}$ (and
the same approximation for $\omega(t)^{-\frac{1}{2}}$)
   then a
simple comparison of eqs.(20)-(21) and (26)-(27) leads to the
conclusion that for large $3H+\tilde{\gamma}^{2}$ the solutions of
the wave equation and the diffusion equation (with zero initial
conditions) coincide.

\section{Power spectrum of the linearized wave equation}
We have calculated the power spectrum in the Einstein-Klein-Gordon
system in \cite{habapl} in the case $U(\phi)=\phi$. The changes
corresponding to the replacement $\phi\rightarrow U(\phi)$ are the
following:$3H+\gamma^{2}\rightarrow 3H+\tilde{\gamma}^{2}$,
$V^{\prime\prime}\rightarrow
V^{\prime\prime}+\frac{1}{2}\partial_{t}\tilde{\gamma}^{2},
\frac{3}{2}\sigma\gamma^{2}H\rightarrow \frac{3}{2}\gamma^{2}H
((U^{\prime})^{2}+UU^{\prime\prime})$. We repeat here the main
steps of \cite{habapl} in order to fix the stage  for the
discussion of the extended model. We still rewrite the
correspondence in a different way. Let
\begin{equation}
\tilde{\Gamma}= (3H)^{-1}\tilde{\gamma}^{2}
\end{equation} replacing  $\Gamma$ from \cite{habapl}, \begin{equation}
Q=\frac{1}{2}\partial_{t}\tilde{\gamma}^{2}+\tilde{\gamma}^{2}(U^{\prime})^{-2}UU^{\prime\prime}
\end{equation}
and
\begin{equation}
\delta=(V^{\prime\prime}+Q)(3H^{2})^{-1}
\end{equation} replacing $\eta$ from \cite{habapl}.
If in eq.(15) we applied the slow roll approximation then we could
express $\partial_{t}\phi_{c}$ in $Q$ by derivatives of the
potential $U$.

 The power spectrum $\rho$ of fluctuations $\phi$ is defined by
\begin{equation} \langle \phi_{t}({\bf x})\phi_{t}({\bf y})\rangle
=\int d{\bf k}\rho_{t}({\bf k})\exp(i{\bf k}({\bf x}-{\bf y}))
\end{equation}or in Fourier  transform
\begin{equation}
\langle \phi_{t}({\bf k})\phi_{t}({\bf
k}^{\prime})\rangle=(2\pi)^{3} \delta({\bf k}+{\bf
k}^{\prime})\rho_{t}({\bf k}).\end{equation} The spectral index
$2\kappa$ is defined by the low $k=\vert {\bf k}\vert$ behaviour
$\rho_{t}({\bf k})\simeq k^{-2\kappa}$.

In \cite{habapl} we have calculated the spectrum under the
assumption that $\delta$ and $\tilde{\Gamma}$ are almost constant.
If $H$ is varying in time then  estimates by means of the methods
\cite{habapl} (based on \cite{bass}) are not reliable for varying
parameters. For this reason in the next section we discuss the
diffusion approximation when the time evolution can be treated in
a more controllable way. We define
 \begin{equation}
 \epsilon=-H^{-2}\partial_{t}H.
 \end{equation}
Without the thermal noise ($\gamma=0$) $\epsilon$  can be
expressed from Friedmann equations as $\frac{1}{8\pi
G}(V^{\prime})^{2}V^{-2}$ (where $G$ is the Newton constant). With
the thermal noise and the interaction $U$ the formula for
$\epsilon$ in terms of potentials is more involved (see
\cite{habaepj},eq.(89)). We keep (33) as a definition of
$\epsilon$ and do not attempt to express it by potentials.

 We introduce the conformal time
\begin{equation}
\tau=\int dt a^{-1}.
\end{equation}With a slowly varying $H$ we have approximately
\begin{equation}
aH=-(1-\epsilon)^{-1}\frac{1}{\tau}.
\end{equation}
Eq.(35) can be obtained by an integration of the identity
\cite{wood}
\begin{equation}
\partial_{t}\Big((1-\epsilon)Ha\Big)^{-1}=-a^{-1}+\partial_{t}\epsilon
\Big(aH(1-\epsilon)^{2}\Big)^{-1}
\end{equation}
and the assumption that the last term on the rhs of eq.(36) is
small in comparison with the first term.

 In terms of $\tau$
eq.(16) for the Fourier transform $\phi({\bf k})$ reads
($k=\vert{\bf k}\vert$)
\begin{equation}\begin{array}{l}
(\partial_{\tau}^{2}-\frac{2+3\tilde{\Gamma}}{1-\epsilon}\frac{1}{\tau}\partial_{\tau}+k^{2}
+\frac{3\delta+\frac{9}{2}\tilde{\Gamma}}{(1-\epsilon)^{2}}\tau^{-2})\phi=\gamma\beta^{-\frac{1}{2}}
\eta_{\tau}.\end{array}
 \end{equation}
Let
\begin{equation}
\zeta=k\tau
\end{equation}
and

\begin{equation}
\nu^{2}=(1-\epsilon)^{-2}\Big(\frac{9}{4}-3\delta-\frac{3}{2}\epsilon+
+\frac{9}{4}\tilde{\Gamma}^{2}-\frac{3}{2}\epsilon\tilde{\Gamma}
\Big)
\end{equation}
(the term $-\frac{3}{2}\epsilon$ in eq.(39) is replaced by
$+\frac{9}{2}\epsilon$ in the corresponding formula in
\cite{habapl} owing to the contribution of gravitational modes as
expressed by scalar perturbations).

The calculation of the expectation value of the solution of
eq.(37) over the noise $\eta$ leads to

\begin{equation}\begin{array}{l}
\langle \phi^{2}\rangle \simeq
k^{-3}\zeta^{2\mu}Y_{\nu}^{2}(\zeta)(U^{\prime}(\phi_{c}))^{2}\simeq
k^{-3}\zeta^{2\mu-2\nu}(U^{\prime}(\phi_{c}))^{2}
\end{array}\end{equation}for small $k$ (as   the Bessel function $Y_{\nu}(\zeta)\simeq \zeta^{-\nu}$ for
 small $\zeta$).Here
\begin{equation}
\mu=(1-\epsilon)^{-1}(\frac{3}{2}-\frac{\epsilon}{2}+\frac{3}{2}\tilde{\Gamma}).
\end{equation}
In eq.(40) the time t in $\phi_{c}(t)$ must be replaced by $\tau$
then $\tau $ is expressed as $\frac{\zeta}{k}$ . For a small
$\tilde{\Gamma}$ and $U^{\prime}\simeq const $ we have in a linear
approximation in the indices describing the interaction
corrections:
\begin{equation}
2\nu=3+2\epsilon -2\delta,
\end{equation}
\begin{equation}
2\mu=3+2\epsilon+3\tilde{\Gamma}.
\end{equation}
From eqs.(40)-(43)  if $U^{\prime}\simeq const$
\begin{equation}
\langle\phi^{2}\rangle\simeq k^{-3+2\delta   +3\tilde{\Gamma}}.
\end{equation}
If  $U(\phi)=\phi$ and the term $\frac{3}{2}\gamma^{2}H\phi$ is
absent in eq.(16) then in \cite{habapl} we obtained the power
spectrum $k^{-3+2\eta}$ which agrees with the corresponding result
in \cite{ram} (in fact, our spectral index $2\mu- 2\nu$ in
\cite{habapl} is equal to $2\nu-2\alpha$ of ref.\cite{ram}). For a
general $U(\phi)$ our wave equation (16) is different from that of
ref.\cite{ram}. The spectral indices in models of warm inflation
in \cite{hal}\cite{mosstaylor}\cite{graham} agree with our results
when they are concerned with spectral indices of the inflaton.
However, there are other fields in those models described by their
entropy and densities which  additionally contribute to the power
spectrum leading to a different spectral index.
\section{The power spectrum of diffusion}

 The solution $\phi_{\eta}=\phi_{c}+\phi$ of eq.(13) with a given initial condition is a sum of the solution $\phi_{c}$
 of the
 homogeneous equation (15) with this initial condition and $\phi$ with 0 as an initial condition at $t_{0}$
 . From the diffusion approximation (26) we obtain
 \begin{equation}\begin{array}{l} \rho_{t}({\bf k})
=\beta^{-1}
\gamma^{2}\int_{t_{0}}^{t}\exp\Big(-2\int_{s}^{t}v\Big)\cr
a(s)^{-3}(3H+\tilde{\gamma}^{2})^{-2}U^{\prime}(\phi_{c})^{2}ds .
\end{array}\end{equation}
We can have $\rho_{t}\simeq k^{-2\kappa}$ with $\kappa >0$ if
$t_{0}=-\infty$ (otherwise the integral (45) would be finite at
$k=0$). For $a=\exp(\int_{t_{0}}^{t}H(t^{\prime})dt^{\prime})$
this means that the initial condition is at $a(t_{0}=-\infty)=0$.
If we introduce the e-fold time
\begin{equation}
d\nu=Hdt,
\end{equation}
then

\begin{equation}\begin{array}{l} \rho_{t}({\bf k})
= \beta^{-1}
\gamma^{2}\cr\int_{\nu(t_{0})}^{\nu(t)}\exp\Big(-2\int_{\tau}^{\nu(t)}
(1+\tilde{\Gamma})^{-1}\cr\Big(\delta+(3H^{2})^{-1}\exp(-2\tau^{\prime})k^{2}
+\frac{3}{2}\tilde{\Gamma} \Big)d\tau^{\prime}\Big) \cr
 \exp(-3\tau)(U^{\prime}(\phi_{c}(\tau))^{2}H^{-1}(3H+\tilde{\gamma}^{2})^{-2}d\tau .
\end{array}\end{equation}
(it is assumed that in $\phi_{c}(s)$ the cosmic time has been
expressed by the e-fold time). We assume in this section that
$\tilde{\Gamma}$, $\delta$ ,$U^{\prime}(\phi_{c})$ , $H$ and
$(U^{\prime})^{-2}U^{\prime\prime}U$ are slowly varying in time,
so that we may approximate them by a constant.

We introduce the variable
\begin{equation}
u=\exp(-2\nu)
\end{equation}
and assume that $H(\nu)\simeq const$ then
\begin{equation}\begin{array}{l}
\rho_{t}({\bf k})=
\frac{1}{2H}\beta^{-1}\gamma^{2}\exp\Big((3H^{2})^{-1}(1+\tilde{\Gamma})^{-1}k^{2}\exp(-2\nu)\Big)
\cr=\exp(-2q\nu)\int^{u(\nu_{0})}_{u(\nu)}(3H+\tilde{\gamma}^{2})^{-2}U^{\prime}(\phi_{c})^{2}
\cr\exp\Big(-(3H^{2})^{-1}(1+\tilde{\Gamma})^{-1}k^{2}u\Big)
u^{\frac{1}{2}-q}du, \end{array}\end{equation}where
\begin{equation}
q=(\delta+\frac{3}{2}\tilde{\Gamma})(1+\tilde{\Gamma})^{-1}.\end{equation}

 The result of
integration in eq.(47) assuming that  $H$,$U^{\prime}$
$\tilde{\Gamma}$ and  $q$  are approximately constant can be
expressed by the incomplete $\Gamma$ function
\begin{equation}\begin{array}{l}\rho_{t}({\bf k})
=\frac{1}{2H}(3H+\tilde{\gamma}^{2})^{-2}\exp(-2q\nu)
\tilde{\gamma}^{2}\beta^{-1}\cr\exp\Big((3H^{2})^{-1}(1+\tilde{\Gamma})^{-1}k^{2}\exp(-2\nu)\Big)\cr
\Big(\Big((3H^{2})^{-1}k^{2}(1+\tilde{\Gamma})^{-1}\Big)^{-\kappa}\cr\Gamma\Big(\kappa,(3H^{2})^{-1}
(1+\tilde{\Gamma})^{-1}k^{2}\exp(-2\nu)\Big) \cr
-\Big((3H^{2})^{-1}(1+\tilde{\Gamma})^{-1}k^{2}\Big)^{-\kappa}\cr\Gamma\Big(\kappa,(3H^{2})^{-1}
(1+\tilde{\Gamma})^{-1}k^{2}\exp(-2\nu_{0})\Big)\Big),\end{array}\end{equation}
where
\begin{equation}
\kappa=\frac{3}{2}-q.
\end{equation}
We have for $ x<<1$
\begin{equation}
\Gamma(\alpha,x)=\Gamma(\alpha)-x^{\alpha}\sum_{n\geq
0}(-x)^{n}\Big(n!(\alpha+n)\Big)^{-1},
\end{equation}
and for  $x>>1$
\begin{displaymath}
\Gamma(\alpha,x)=x^{\alpha -1}\exp(-x).
\end{displaymath}
If $\nu_{0}\rightarrow -\infty (u(\nu_{0})\rightarrow +\infty)$
then the second term in eq.(51) is vanishing. There remains
\begin{equation}\begin{array}{l}\rho_{t}({\bf k})
=\frac{1}{2H}(3H+\tilde{\gamma}^{2})^{-2}\exp(-2q\nu)
\tilde{\gamma}^{2}\beta^{-1}\cr\exp\Big((3H^{2})^{-1}(1+\tilde{\Gamma})^{-1}k^{2}\exp(-2\nu)\Big)
\cr\Gamma\Big(\kappa,(3H^{2})^{-1}
(1+\tilde{\Gamma})^{-1}k^{2}\exp(-2\nu)\Big)\cr\Big((3H^{2})^{-1}k^{2}(1+\tilde{\Gamma})^{-1}\Big)^{-\kappa}
.\end{array}\end{equation} From eq.(53)only the last term in
eq.(54) is relevant for a small $k$ leading to
\begin{equation}\begin{array}{l}\rho_{t}({\bf k}),
\simeq k^{-2\kappa}.\end{array}\end{equation} where in a linear
approximation in the indices $\delta$ and $\tilde{\Gamma}$ we get
\begin{equation}
2\kappa=3-2\delta-3\tilde{\Gamma}
\end{equation}
 This result agrees with the result (44) obtained from the wave
 equation in sec.4. For a large $\Gamma$ eqs.(40) and (55) with
 $\kappa$ defined in eq.(52) also give the same results but power spectrum
 is far from the scale invariant one in contradistinction to the
 models in \cite{hal} (but in an agreement with the calculations of
 the spectrum of the inflaton stochastic equation in \cite{ram}).

At $\gamma=0$ the result (56) coincides with the power spectrum of
quantum fluctuations which are derived  by a calculation of
$\langle\phi^{2}\rangle$ in the Bunch-Davis vacuum
\cite{lin1}\cite{vil}(normalized so that the scalar modes behave
as plane waves at large $k(aH)^{-1}$; see also a later discussion
in \cite{linde}\cite{abbott}\cite{tak}\cite{lyt}(sec.24.3)). Our
results agree with the results of \cite{ram}(also with the
calculations of \cite{hal}\cite{graham} when the authors calculate
the power spectrum of the stochastic equations). In comparison
with \cite{ram} one should take into account that we have a
different friction term than the authors in\cite{ram} and the term
$\frac{3}{2}\gamma^{2}H(U^{\prime 2}UU^{\prime\prime}$) is absent
in \cite{ram}) . For $U(\phi)=\phi$ we have calculated the
spectrum of the same model as in \cite{ram} in \cite{habapl}(we
consider there an extra term of the form
$\sigma\frac{3}{2}\gamma^{2}H\phi$  which for  $\sigma=0$
corresponds to the case of \cite{ram}). Then, our results agree.
For general $U(\phi)$ the spectrum of the inflaton equation of
\cite{ram} cannot be compared with eq.(16) without additional
calculations. It follows from eq.(54) that the amplitude of
thermal fluctuations
 is determined by $H$,
$\kappa$ (known from CMB measurements \cite{lin2}\cite{lin}),
$\beta$ and $\gamma$ (which this way would be fixed by
$\rho_{t}({\bf k})$). On the other hand the friction $\gamma$ is
related (depending on the model) to other measurable quantities as
,e.g., the diffusion constant  \cite{habasz}. In this way the
amplitude of stochastic thermal fluctuations depends on many
parameters, whereas the virtue of the quantum result consists in
the prediction of its $10^{-5}$ magnitude
\cite{muk}\cite{guthpi}\cite{starjetpl}\cite{starjpl}\cite{linde}
in agreement with observations.  The theory shows that under the
assumption of almost exponential expansion both the quantum
fluctuations and the thermal fluctuations of the inflaton lead to
a small deviation from the scale invariant spectral index ( this
index is crucial
 for distinguishing various inflation models on the basis of observational data
\cite{lin2} \cite{lin}). The assumption that thermal fluctuations
are of quantum origin does not change essentially the results as
 at high temperatures at the early stage of the universe quantum theory is well approximated by the classical
 one.
 Although CMB shows the quantum Planck spectrum (at all wave
 lengths) the perturbations of the homogeneous solutions at large
 wave lengths exhibit no quantum effects.

 In the next section we show that the potential $U$ describing
 an interaction of the inflaton  with the environment can shift the spectral index.
 It may be difficult on the basis of a study of the power spectrum to
 determine whether the deviation from the scale invariant spectrum
 discovered in WAMP observations comes from quantum  or
  thermal fluctuations. If the initial state of the universe  is Gaussian then
  further evolution of quantum cosmological perturbations proceeds in a squeezed state
  with a classical
  evolution as shown in \cite{kiefer0}\cite{kiefer}\cite{kiefer2}. In such a case it would be
  difficult to discover whether the origin  of
  the universe is of quantum nature. The eventual observation of
  non-Gaussian correlations \cite{maldacena} in CMB could show that a decoherence of quantum
  superpositions really takes palace.

\section{Beyond the slowly varying corrections}
In the calculations of the spectrum of the stochastic wave
equation in sec.4 as well as of the spectrum of the diffusion
equation in sec.5 we assumed that $H$,$\delta$ and
$\tilde{\Gamma}$ vary so slowly that we can approximate them by
constants in the calculation of the power spectrum. The slow
variation is consistent with the slow roll approximation usually
made for inflation. The calculations in sec.4 relied heavily on
the assumption of the slow variation. For varying potentials we
return to the approximation of sec.3 of the wave equation by the
diffusion equation. The replacement of the wave equation by
diffusion equation is legitimate if
$3H+\gamma^{2}(U^{\prime})^{2}$ is large. The estimates of the
solutions of the diffusion equation based on eq.(26) are much
easier then the study of the corresponding wave equation. We
rewrite the formula (45) for the spectrum in the form
\begin{equation}\begin{array}{l} \rho_{t}({\bf k})
= = \beta^{-1}
\gamma^{2}\cr\int_{t_{0}}^{t}\exp\Big(-2\int_{s}^{t}(3H(t^{\prime})
+\tilde{\gamma}^{2})^{-1}\cr\Big(V^{\prime\prime}+a^{-2}k^{2}
+\frac{1}{2}\partial_{t}^{\prime}\tilde{\gamma}^{2}+\frac{3}{2}\gamma^{2}H((U^{\prime})^{2}+U^{\prime\prime}U)
\Big) \cr
 a(s)^{-3}(3H+\gamma^{2}U^{\prime}(\phi_{c}(s))^{2})^{-2}(U^{\prime}(\phi_{c}(s))^{2}ds .
\end{array}\end{equation} In this formula we admit that
$V^{\prime\prime}(\phi_{c}(t))$ and $U^{\prime}(\phi_{c}(t))$ have
a substantial variation in time. This property depends on the
potentials as well as on $H$. First, we assume that $H$ is
(almost) constant. In the damped wave equation the large time
behaviour does not depend on the term $\partial_{t}^{2}\phi$. For
small time in order to neglect $\partial_{t}^{2}\phi$ we must make
the assumptions that $(V^{\prime}V^{-1})^{2}$ and
$V^{\prime\prime}V^{-1}$ are small and slowly varying in time
(this is the  slow-roll  approximation). Neglecting
$\partial_{t}^{2}\phi_{c}$ in eq.(15) we can represent this
equation in an integral form
\begin{equation}
\int(V^{\prime}(\phi_{c})
+\frac{3}{2}\gamma^{2}HUU^{\prime}(\phi_{c}))^{-1}(3H+\gamma^{2}U^{\prime}(\phi_{c})^{2})d\phi_{c}=-t
\end{equation}
Eq.(15) will have solutions decaying to zero (or to a constant) as
from eq.(58) $\partial_{t}\phi_{c}$ is negative if $ V^{\prime}
+\frac{3}{2}\gamma^{2}UU^{\prime}$ is positive. If the decay is
exponential
\begin{equation}
\phi_{c}\simeq \exp(-b t)
\end{equation}
so that \begin{equation}U^{\prime}(\phi_{c}(t))\simeq \exp(-r t)
\end{equation}
then, as follows from the estimates of sec.5 after an insertion of
(60) in eq.(57)
\begin{equation}
\rho_{t}(k)\simeq k^{-3-2r+2\delta+3\tilde{\Gamma}}.
\end{equation}Hence, the decay (60) leads to a shift of the
spectral index. The behaviour (61) can really happen as we can see
assuming that $V^{\prime}$ is negligible  and
$\gamma^{2}(U^{\prime})^{2}>>3H$. Then, eq.(58) has the solution
\begin{equation}
U(\phi_{c}(t))=A\exp(-\frac{3H}{2}t).
\end{equation}
If $U\simeq \phi^{n} $ then  $r=\frac{3H(n-1)}{2n}$. An
exponential decay will be a common behaviour for polynomial $V$
and $U$ in eq.(58).  Let us consider some examples.
$V=\frac{m^{2}}{2}\phi^{2}$, $U=\phi$ gives a linear equation (15)
with the decay rate
$b=(m^{2}+\frac{3}{2}\gamma^{2}H)(3H+\gamma^{2})^{-1}$. Easily
calculated integral (58) for $V=\frac{m^{2}}{2}\phi^{2}$ and
$U=\frac{1}{2}\phi^{2}$ gives \begin{equation}
b=r=\frac{m^{2}}{3K}.
\end{equation}
As a next example if $V=\frac{g}{4}\phi^{4}$, $U=\phi$ then
$b=\frac{1}{2}\gamma^{2}$, but $r=0$ (no effect on the power
spectrum in eq.(61)). The decay can be non-exponential as can be
seen if $V=\frac{g}{4}\phi^{4}$ and $U=\frac{1}{2}\phi^{2}$ then
\begin{displaymath}
\phi_{c}^{-2}=\phi_{0}^{-2}+\frac{4g+3H\gamma^{2}}{6H}t.
\end{displaymath}
In such a case $b=r=0$ and the power spectrum is changed only by
logarithmic corrections.

As a different class of models let us consider the power-law
inflation $a=t^{\alpha}$ , $H=\frac{\alpha}{t}$. Consider the
potentials
\begin{equation}
V(\phi)=\lambda\exp(4u\phi)
\end{equation}
and
\begin{equation}
U(\phi)=\Lambda\exp(u\phi).
\end{equation}
Eq.(15) has a solution of the form \begin{equation}
\phi=-\frac{1}{2u}\ln(t)
\end{equation}
if the parameters satisfy the relation
\begin{equation}
\alpha=\frac{1}{3}\frac{1-u^{2}\gamma^{2}\Lambda^{2}+8\lambda
u^{2}}{1-u^{2}\gamma^{2}\Lambda^{2}}.
\end{equation}
We require $\alpha>1$, hence $u^{2}\gamma^{2}\Lambda^{2}<1$.

The $s$-integral  in eq.(57) reads (we choose $t_{0}=0$ so that
$a(t_{0})=0$ )
\begin{equation}
\rho_{t}=K(t,k)\int_{0}^{t}ds s^{-3\alpha+h+1} \exp\Big(
-B(2\alpha-2)^{-1}k^{2}s^{2-2\alpha}\Big),
\end{equation}
where $K$ is a certain function bounded for a small $k$,

\begin{equation}\begin{array}{l}
h=2(3\alpha+\gamma^{2}\Lambda^{2}u^{2})^{-1}\Big(16\lambda u^{2}
-\frac{1}{2}\gamma^{2}\Lambda^{2}u^{2} \cr+
\gamma^{2}\Lambda^{2}u^{2}\frac{1+8\lambda
u^{2}-\gamma^{2}\Lambda^{2}u^{2}}{1-\gamma^{2}\Lambda^{2}u^{2}}\Big),
\end{array}\end{equation}
\begin{equation}
B=2(3\alpha+\gamma^{2}\Lambda^{2}u^{2})^{-1}.
\end{equation}
Performing the integral (68) we obtain the power spectrum (55)
with
\begin{equation}
\kappa=\frac{3\alpha-h}{2(\alpha-1)}.
\end{equation}
For $h=1$ the result is the same as in the case of a power
spectrum of quantum fields \cite{abbott} in a metric $a\simeq
t^{\alpha}$ (this is the almost scale invariant spectrum for a
large $\alpha$)

\section{Summary and outlook}
The stochastic wave equation can be considered as a
phenomenological effective field theory of an inflaton. In
general, in addition to the inflaton there will be other fields
which contribute to the power spectrum. In our model the
stochastic wave equation arises from an average over an infinite
set of scalar fields interacting with the inflaton. We have
investigated its long wave power spectrum on a basis of some
well-controlled approximations. We have shown that the scale
invariant spectrum is related to the coordinate-independent form
of the noise and to the accelerated expansion of the metric. The
diffusion approximation derived in this paper allows to study the
inflaton power spectrum beyond the assumption of an almost
exponential expansion and small variation of the potentials. We
considered a potential $U$ describing interaction with an
environment which in the case of an almost exponential expansion
shifted the spectral index. In an example of a power-law inflation
we have obtained a power spectrum close to the scale invariant one
in models with exponential potentials.
  For a small friction
and an  almost exponential expansion the departure from the scale
invariant spectrum is determined by the same formula as the one
obtained from quantization of the scalar field on an external
expanding space-time. If there is a friction then Hamiltonian
quantum mechanics is not well-defined. However, we suppose that
the proper formulation as a dissipative Lindblad theory would lead
the same formula for the power spectral index. Our stochastic
methods suggest that the spectral long wave index cannot
distinguish between quantum inflaton fluctuations and classical
thermal fluctuations. The time evolution of cosmological
perturbations has been studied in \cite{kiefer0}\cite{kiefer} with
the conclusion that if the inflation starts from a Gaussian state
then it quickly becomes classical (decoherence without the
environment). The CMB spectrum satisfying the Planck law is
certainly quantum but we could not see this in the long wave
limit. If we admitted non-Gaussian states then a complete
decoherence theory based on the  Lindbald equation would be needed
\cite{kiefer2} in order to explain the structure formation and
detect when the classical behaviour begins. From the formula for
$\rho_{t}(k)$ in this paper we could conclude that
$\rho_{t}(k)\simeq k^{-1}$ for large $k$. This is a quantum
behaviour of $\langle\phi^{2}\rangle $. However, for large $k$ the
stochastic equation discussed in this paper is not reliable. One
should rather study the interaction with an environment at high
momenta initiated in \cite{habaepj}.
\section{Appendix A:Invariance under a change of coordinates} We
give a simple proof that the stochastic wave equation without
friction (friction comes from an interaction of $\phi$ with  an
environment as in \cite{habaepj}, the low momentum approximation
is not invariant under change of coordinates)
\begin{displaymath}
g^{-\frac{1}{2}}\partial_{\mu}g^{\frac{1}{2}}\partial^{\mu}\phi
+V^{\prime}=g^{-\frac{1}{4}}\eta \end{displaymath}is invariant
under a change of coordinates, where
$g=\vert\det[g_{\mu\nu}]\vert$. Under the change of coordinates
$x\rightarrow y$
\begin{equation}\begin{array}{l}
\langle \eta(x)\eta(x^{\prime})\rangle= \delta(x-x^{\prime})
=\delta\Big(x(y)-x^{\prime}(y^{\prime})\Big)\cr =
\delta(y-y^{\prime})\vert \frac{\partial x}{\partial y}\vert^{-1}
= \vert \frac{\partial x}{\partial y}\vert^{-1}\langle
\eta(y)\eta(y^{\prime})\rangle
\end{array}\end{equation} where $\frac{\partial x}{\partial y}\vert$ is the Jacobian.So
\begin{equation}
\eta(x)=  \vert \frac{\partial x}{\partial
y}\vert^{-\frac{1}{2}}\eta(y)
\end{equation}
On the other hand
\begin{equation}
g(x)=g(y)\vert \frac{\partial x}{\partial y}\vert^{2}
\end{equation}
Hence, $g^{-\frac{1}{4}}\eta $ is invariant (in a flat expanding
metric this is $a^{-\frac{3}{2}}\eta$).
 \section{Appendix B:Exact
formula for the exponential expansion} When $a(t)=\exp(Ht)$ then
the solution of the diffusion can be obtained explicitly. Assume
$U(\phi)=\phi$, denote
\begin{equation}
M^{2}=V^{\prime\prime}+\frac{3}{2}\tilde{\gamma}^{2}.
\end{equation}and assume that $M^{2}$ can be approximated by a
constant. The solution of the linear diffusion equation (19) with
zero initial condition at  $t_{0}=-\infty $ ( $a(-\infty)=0$) is
\begin{equation}\begin{array}{l}
\phi_{t}=\beta^{-\frac{1}{2}}\tilde{\gamma}\int_{-\infty}^{t}ds\frac{1}{3H+\tilde{\gamma}^{2}}\eta_{s}\exp(-\frac{3}{2}H(t-s))
\cr\exp\Big(-\frac{k^{2}}{3H+\tilde{\gamma}^{2}}
(\exp(-2Hs)-\exp(-2Ht))\cr-\frac{M^{2}}{3H^{2}+H\tilde{\gamma}^{2}}(t-s)
\Big)\end{array}
\end{equation}
Let
\begin{displaymath}
u(s)=\exp(-2Hs)
\end{displaymath}
\begin{equation}
R=3H^{2}(1+\frac{1}{3}\tilde{\gamma}^{2}H^{-1})=3H^{2}(1+\tilde{\Gamma})
\end{equation}
Then\begin{equation}\begin{array}{l} \rho_{t}({\bf
k})=\frac{\tilde{\gamma}^{2}}{2H\beta}(\frac{1}{3H+\tilde{\gamma}^{2}})^{2}
\exp(\frac{k^{2}\exp(-2Ht)}{R})
\cr\int_{u(t)}^{\infty}\exp(-\frac{k^{2}u}{R}) u^{\kappa-1}du
\end{array}\end{equation} where
\begin{equation}
\kappa=\frac{3}{2}-\frac{M^{2}}{R}
\end{equation}The integral can be expressed by the incomplete
$\Gamma$
\begin{equation}\begin{array}{l}
 \rho_{t}({\bf
k})=\frac{\tilde{\gamma}^{2}}{2H\beta}
(\frac{1}{3H+\tilde{\gamma}^{2}})^{2}\exp(\frac{k^{2}\exp(-2Ht)}{R})
\cr(\frac{k^{2}}{R})^{-\kappa}
\Gamma(\kappa,\frac{k^{2}}{R}u(t)).\end{array}\end{equation} From
eqs.(53) and (80)\begin{equation} \rho_{t}({\bf k})\simeq
k^{-2\kappa}
\end{equation}
 If $\gamma=0$ then
\begin{equation} \kappa =\frac{3}{2}-\delta
\end{equation}with $\delta=\frac{M^{2}}{3H^{2}}$ .

This is exactly the index resulting from a quantization of  the
scalar field in an exponentially expanding universe
\cite{linde}\cite{abbott}\cite{tak}\cite{lyt}.


\begin{thebibliography}{99}
\bibitem{har}E. Harrison, Phys.Rev.D{\bf 1},2726(1970)
\bibitem{zel}Ya.B.Zeldovich, MNRAS,{\bf 160},1(1972)


\bibitem {sto}K.T. Story  et al(2013) ApJ,  779,86
\bibitem{ade}  P.A.R. Ade  et al (2015) arXiv:1502.01589
 \bibitem{starjetpl}A.A. Starobinsky, JETP Lett.{\bf
30},682(1979)
\bibitem{muk}V.F.Mukhanov ,G.V. Chibisov  (1981) JETP
Lett. 33,532
\bibitem{starjpl}A.A. Starobinsky, Phys.Lett.{\bf 117B},175(1982)
\bibitem{lin1}A.D. Linde,Phys.Lett.{\bf B116}335(1982)
\bibitem{vil}A. Vilenkin and L.H. Ford, Phys.Rev.{\bf
D26},1231(1982)
\bibitem{guthpi}A.H. Guth and S.-Y. Pi, Phys.Rev.Lett.{\bf
49},1110(1982)\bibitem{linde} A.O. Linde, JETP Lett.{\bf
40},1335(1984)


\bibitem{bardeen} J. Bardeen, Phys.Rev.D{\bf 22},202(1980)
 \bibitem{feld}V.F. Mukhanov,H.A. Feldman,
 R.H.Brandenberger,

 Phys.Rep.{\bf 215},203(1992)
 \bibitem{starjettpl}A. A. Starobinsky, JETP Lett.{\bf
 42},152(1985)
\bibitem{marozzi}F. Finelli, G. Marozzi G.P. Vacca and G. Venturi,


Phys.Rev.{\bf D69},123508(2004)
\bibitem{marozzi2}F. Finelli, G. Marozzi , A.A. Starobinsky, G. Venturi,


Phys.Rev.{\bf D79},044007(2009)


\bibitem{starven} V. Vennin, A.A. Starobinsky, Eur.Phys.J.C
{\bf 75},413(2015)


 \bibitem{hal} M.H. Hall, I.G. Moss  ,A. Berera  ,


 Phys.Rev.
    D{\bf 69},083525(2004)
 \bibitem{ram} R.O. Ramos and L.A. da Silva   JCAP03,032(2013)


\bibitem{berera2}A. Berera, Nucl.Phys.B {\bf 585},666(2000)
\bibitem{bererarev}A. Berera ,I.G. Moss and R.O. Ramos ,  Rep.Progr.Phys.
72,026901(2009)
\bibitem{mosstaylor}A.N. Taylor and A. Berera, Phys.Rev.{\bf
D62},083517(2000)
\bibitem{graham}Ch. Graham and I.G. Moss,

JCAP7,013(2009) ,arXiv:0905.3500









 \bibitem{ber1} A. Berera Phys.Rev. D{\bf 54},2519(1996)
\bibitem{hab1}Z. Haba,Adv.High.Energy.Phys.


    2018,7204952(2018);arXiv:1807.00639

\bibitem{habaepj}Z. Haba, Eur.Phys.J. C{\bf
80},321(2020);arXiv:2001.07268(gr-qc)

 \bibitem{bass}B. A. Bassett ,  S. Tsujikawa and
D. Wands

 Rev.Mod.Phys.78,537(2006)

\bibitem{habapl}Z.Haba, Phys.Lett.B {\bf 792},320(2019)




\bibitem{wood} R.P. Woodard, Int.Journ. Mod.Phys.{\bf
D23}1430020(2014)

\bibitem{abbott} L.F.Abbott and M.B.Wise  , Nucl.Phys.  B{\bf 244},541(1984)
 \bibitem{tak}F.Takahashi , Wen Yin, A.H. Guth,


Phys.Rev. D{\bf 98},015042(2018)

 \bibitem{lyt}D.H. Lyth and A.R. Liddle  The Primordial
Density Perturbation,

 Cambridge University Press, Cambridge,2009




\bibitem{indian} K. Bhattacharya, S. Mohanty,
R.Rangarajan,

Phys.Rev.Lett.{\bf 96},121302(2006)




\bibitem{lin2}E.J.Copeland,E.W. Kolb   A.R. Liddle, J.E.Lindsey ,


Phys.Rev.Lett.{\bf 71},219(1993)

 \bibitem{lin} E.J. Copeland, J.E. Lindsey  ,  A.R. Liddle , E.W. Kolb,

 T.  Barreiro and
M. Abney, Rev.Mod.Phys. 69,374(1997)



\bibitem{habasz}Z.Haba,A. Stachowski and M. Szydlowski ,

JCAP07(2016)024













\bibitem{kiefer0}A. Albrecht, P.Ferreira, M. Joyce, T.Prokopec,


Phys.Rev.{\bf D50},4807 (1994)
\bibitem{kiefer}D. Polarski, A.A. Starobinsky,

Class.Quantum Gravity,{\bf 13},377(1996)
\bibitem{kiefer2}C. Kiefer, D. Polarski,

 Adv.Sci.Lett.{\bf
2},164(2009), arXiv:0810.0087
\bibitem{maldacena} J. Maldacena, JHEP 05(2003)013
\end{thebibliography}
\end{document}